# Accurate six-band nearest-neighbor tight-binding model for the $\pi$-bands of bulk graphene and graphene nanoribbons

Timothy B. Boykin[a], Mathieu Luisier[b], Gerhard Klimeck[b], Xueping Jiang[c], Neerav Kharche[c], Yu Zhou[c], and Saroj K. Nayak[c]

[a]*Department of Electrical and Computer Engineering, The University of Alabama in Huntsville, Huntsville, Alabama 35899 USA*

[b]*Network for Computational Nanotechnology, School of Electrical and Computer Engineering, Purdue University, West Lafayette, Indiana 47907 USA*

[c]*Department of Physics, Applied Physics, and Astronomy, Rensselaer Polytechnic Institute, Troy, New York 12180 USA*

Accurate modeling of the $\pi$-bands of armchair graphene nanoribbons (AGNRs) requires correctly reproducing asymmetries in the bulk graphene bands as well as providing a realistic model for hydrogen passivation of the edge atoms. The commonly used single-$p_z$ orbital approach fails on both these counts. To overcome these failures we introduce a nearest-neighbor, three orbital per atom $p/d$ tight-binding model for graphene. The parameters of the model are fit to first-principles density-functional theory (DFT) – based calculations as well as to those based on the many-body Green's function and screened-exchange (GW) formalism, giving excellent agreement with the *ab initio* AGNR bands. We employ this model to calculate the current-voltage characteristics of an AGNR MOSFET and the conductance of rough-edge AGNRs, finding significant differences versus the single-$p_z$ model. These results show that an accurate bandstructure model is essential for predicting the performance of graphene-based nanodevices.

## I. INTRODUCTION

Since the first experimental demonstration of monlayer graphene structures[1-3], their unusual quasi-linear band-dispersion and high bulk mobility[4,5] have attracted much attention as potential candidates to augment or replace Si as the material for next-generation nanotransistors. However, bulk graphene has no band gap, making it unsuitable for logic applications. On the other hand, graphene nanoribbons (sheets less than 10nm wide) can have noticeable band gaps, thus becoming semiconducting devices[6-9]. Particularly strong candidates for next-generation nanodevices are nanoribbons in the armchair configuration (AGNRs) because their use in field-effect transistors is expected to lead to improved ON- and OFF-currents.

To date most modeling of the $\pi$-bands of graphene has been carried out with the single-$p_z$ orbital model introduced over sixty years ago by Wallace.[10] Its widespread use for both bulk and a variety of nanostructures[11-14] is doubtless due to its simplicity and computational efficiency, especially for transport simulations. However, the model's virtue, its simplicity, raises a significant question concerning its use for nanostructure simulations: Does it include sufficient physical content to accurately calculate AGNR bandgaps and shapes? An affirmative answer to this question requires the model: (i) to accurately reproduce the *ab initio* bulk graphene bands in the region around $K$ and along $K$-$M$ from which the major components of the AGNR bands come; and (ii) to accommodate a realistic hydrogen passivation approach. We show here that the answer to this question is negative on both counts.

First, the single-$p_z$ model cannot reproduce the asymmetry at $M$ found in *ab initio* calculations, as shown in Fig. 1. Here we plot bulk graphene bands as calculated with three different approaches: Density-functional theory with GW corrections (DFT+GW) (diamonds), the single-$p_z$ model (dotted lines) and a $p/d$ model to be introduced below (solid lines). Note the error of around 0.8 eV in the single-$p_z$ gap at $M$ as compared to the other two models. This error is important because states in this part of the bulk bandstructure contribute strongly to the central AGNR conduction and valence bands.

Second, and more seriously, the single-$p_z$ model does not allow for any realistic hydrogen passivation approach. If the hydrogen atoms are modeled with only the ground state (a single $s$-orbital), then there is absolutely no coupling to the $\pi$-bands and hence no passivation. It is for this reason that most AGNR calculations using the single-$p_z$ model are for unpassivated structures. As a result, AGNRs of the $3n+2$ family have zero gap in the single-$p_z$ model, contrary to recent first-principles calculations.[15] This incorrect behavior is shown in Fig. 2(a), which graphs the AGNR bandgaps as calculated with DFT-LDA (solid diamonds), the $p/d$ model (open circles), and the single-$p_z$ model (open squares). In addition to the consistent zero-gap result for the $3n+2$ family, the single-$p_z$ approach exhibits almost identical $3n$ and $3n+1$ gap curves and a generally poor agreement with DFT for all three families. Including a single $p_z$ orbital on the passivating hydrogens does not improve the situation, because the single set of passivation parameters $\left(V_{pHpC\pi}, E_{pH}\right)$ has not proven able to fit the DFT gaps of all three AGNR families. Typically, a set of passivation parameters which succeeds with one family fails for another. These shortcomings demonstrate that the single-$p_z$ model is just too simplistic to accurately model AGNR nanotransistors and other structures, rendering it useless for performance comparisons of AGNR nanotransistors to conventional Si MOSFETs. Therefore, a better approach is needed.

An improved graphene model must therefore overcome both of these critical failures of the single-$p_z$ approach, but to be useful for nanodevice simulations must be structured to efficiently interface with transport calculations. DFT certainly incorporates sufficient physics, but at an unacceptably high computational cost. To date it has only been used in thin AGNR transport calculations[16,17], and in any event is still too intensive for iterated design cycles. Non-orthogonal tight-binding[18] and a third-nearest-neighbor $\pi$-bonded model[19] have been proposed to address the poor bulk reproduction of the single-$p_z$ model, however the lack of orthogonality and more-distant neighbor interactions both reduce the efficiency for transport calculations and make the programming aspects of interfacing to nearest-neighbor Si tight-binding models commonly used in device simulations problematic[20].

Our solution to this dilemma is a relatively simple, nearest-neighbor $\left\{ p_z, d_{yz}, d_{zx} \right\}$ orthogonal tight-binding model which includes the essential physics of both bulk graphene and hydrogen passivation as used in AGNRs. The model is computationally efficient and interfaces well with multi-band nearest-neighbor transport models. For accuracy, we parameterize the model using density-functional theory calculations (DFT+GW for bulk and DFT-LDA for passivation parameters in AGNRs). This parameterization thus allows device simulations to be performed with the accuracy of DFT, but at a computational burden not much larger than the single-$p_z$ model. Below we first briefly introduce the model, then discuss our passivation approach and the resulting

AGNR bandgap calculations. Finally we compare device characteristics calculated with both the *p/d* and single-$p_z$ models. In particular, we examine the current-voltage characteristics of the AGNR-MOSFET (metal-oxide semiconductor field-effect transistor) studied by Fiori, et. al[14], as well as the differential conductance of two different types of rough AGNRs.

## II. BULK AND HYDROGEN PASSIVATION MODELS

We choose a three-orbital per atom nearest-neighbor basis $\{p_z, d_{yz}, d_{zx}\}$ because it is the simplest model capable of accurately reproducing DFT+GW bulk graphene bands in the region around *K* and especially along *K-M*; the LDA bulk graphene bands do not differ radically from the GW results. The nearest-neighbor $p_z$-only (or $\pi$-bonded) model is not accurate in this region, since its conduction- and valence-bands at *M* are perfectly symmetric about *K*, as shown in the Appendix. Adding more orbitals, such as an excited $p_z$ (suggested by the structure of atomic carbon) does not greatly improve the accuracy on *K-M*. In our experience, it seems essential to have *d*-orbitals in the basis set. Because the orbitals $\{s, p_x, p_y, d_{xy}, d_{x^2-y^2}, d_{3z^2-r^2}\}$ do not couple to our set for perfectly flat graphene, they are omitted. We employ the same basis set for passivating hydrogen atoms as well. While this choice may seem unusual, it should be recalled that the $n = 2$ and $n = 3$ levels of atomic hydrogen are separated by less than 2eV, and furthermore, if only the 1*s*-orbital is included, the AGNR $\pi$-bands cannot be passivated. The passivation model is discussed in detail below.

Our bulk graphene parameters are listed in the left-hand part of Table I, and the DFT+GW bands used to fit them are plotted in Fig. 1(a). Our first principles calculations are based on local spin density functional theory (DFT) and wave functions are expanded in terms of plane waves. In particular we used Trouiller-Martins norm-conserving pseudopotentials with 30 Hatree cutoffs. The calculations were based on super-cell approach and the interlayer distance used was large enough to minimize the interaction between periodic images. The ground state electronic properties were first obtained based on local density approximation (LDA) to the DFT and quasi-particle gaps were computed based on the GW scheme. Full GW corrections were computed at $\Gamma$, $K$, and $M$ as well as at the midpoints of the symmetry lines connecting them. Polynomial interpolation was used for GW corrections at other points.

The bulk graphene bands reproduced by our model are graphed in the vicinity of the $K$-point in Figure 1(b) (solid lines), along with the DFT+GW bands (diamonds) and the conventional $p_z$-only bands for $E_p = 0.12742\text{eV}, V_{pp\pi} = -2.7\text{eV}$ (dotted lines). As a reference the $K$-points of all three models are aligned. The $p/d$ model gives superior reproduction of the DFT+GW bands, including the asymmetry at $M$; we demonstrate this property with analytic expressions in the Appendix. We emphasize that regardless of the parameters, the nearest-neighbor $p_z$-only model cannot reproduce this asymmetry.

As is the case with bulk graphene, our $p/d$ model permits a more realistic treatment of AGNR bands, both in terms of the gaps and the overall bandstructures themselves. An important part of this improvement over the $p_z$-only model is the better handling of hydrogen passivation. We tried modeling the passivating hydrogen with a single $p_z$- orbital in both the $p_z$-only and our *p/d* model for graphene. In neither case did we achieve acceptable results. In the $p_z$-only model making the hydrogen-carbon nearest-neighbor matrix element sufficiently large to open a significant gap for the $3n+2$ family of AGNRs tended to result in highly inaccurate gaps for the other families; similar behavior occurred in the *p/d* model. (We adopt the convention of Ref. 15 for AGNR indices.) In the *p/d* model this deficiency likely arises from the abrupt termination of the two *d*-components of the wavefunction when only a single $p_z$-orbital is used for hydrogen. In the $p_z$-only model the hydrogens were effectively acting as extra carbons, artificially lengthening the AGNR; the single H-C coupling parameter proved ineffective at softly terminating the wavefunction. Consequently, we include the full orbital set for hydrogen in the *p/d* model. The parameters are listed in the right-hand half of Table I and were optimized to DFT calculations for *only* the trio AGNR-7, -8, -9, with all carbon atoms in their ideal positions and the H-C bond angle identical to the C-C bond angle. For all other AGNRs we employ this same set of parameters and achieve very good agreement between our tight-binding model and DFT for all AGNR families.

At this juncture some brief remarks on the differences between the AGNR bands as calculated with and without GW corrections are in order. Our own GW calculations (not shown) are in line with those of Ref. 15, predicting significantly larger bandgaps for all AGNRs as compared to the LDA results presented here. Enhanced Coulomb effects are the likely cause of the difference. Compared to bulk, AGNRs in vacuum are under confinement and have greatly reduced screening, both of which magnify Coulomb effects. The enhanced interactions appear to so grossly distort the electronic structure that an essentially bulk-like description is no longer valid[15]. More specifically, it appears that the GW-corrected AGNR bands cannot be represented by the relatively small subset of bulk graphene bands included in even the *p/d* model, to say nothing of the $p_z$-only model. The relevant question for device simulation, though, is whether the conditions leading to the large GW corrections accurately describe the environment typical of AGNR nanotransistors. Because AGNRs in nanotransistors are generally surrounded by high-k dielectrics, we believe that the Coulomb effects in them will be less than those in AGNRs in vacuum. Thus we fit our passivation parameters to the LDA AGNR bandgaps.

Figure 2(a) shows the gaps reproduced by DFT (diamonds), the *p/d* model (circles), and the $p_z$-only model (squares) for the three families of AGNRs: $3n$ (solid lines), $3n+1$ (dashed lines), and $3n+2$ (dotted lines). The DFT (without GW corrections) and *p/d* model AGNRs are hydrogen passivated while the $p_z$-only model AGNRs have no passivation and $V_{pp\pi} = -2.7\text{eV}$, as discussed above. There is excellent agreement between the *p/d* model and the DFT calculations, while the $p_z$-only model fails to agree well for any family of AGNRs. Note in particular that in the $p_z$-only model the $3n$ and $3n+1$ results are nearly the same for adjacent AGNRs, while the $3n+2$ family consistently and incorrectly predicts a zero gap.

Figures 2(b)-(d) show the bands in the first half of the Brillouin zone for the AGNR series -11, -12, -13. We emphasize that none of these three AGNRs was used to optimize the H-C parameters. Because there is no absolute energy in DFT, we align the DFT uppermost valence band edge with the *p/d* model for each AGNR. To facilitate

comparison with the other two models we also align the uppermost valence band of the $p_z$-only model with the $p/d$ model, although in reality its bands should be shifted relative to the $p/d$ bands since the bulk $K$-points of the two models are aligned. Note that the $p_z$-only results miss the DFT results by significant margins, in terms of both band gaps and band shapes. On the other hand, using the *same* passivation parameters for *all* AGNRs, the $p/d$ model results agree very well with the DFT calculations, both in the gaps and the overall band shapes and positions. We emphasize that the gaps in the DFT and $p/d$ models are not due to edge disorder: All carbons are in ideal positions and the H-C bond angle is the same as the C-C angle.

The $p_z$-only model has three further anomalies. One is that for AGNRs with odd indices are perfectly flat bands at $E_p \pm V_{pp\pi}$ (not shown here). This behavior is not a numerical artifact, but is instead an inherent deficiency of the model: it has been observed in the analytical $p_z$-only AGNR calculations of Ref. 12. Second, the $p_z$-only model introduces an artificial symmetry for all AGNRs (only odd-index AGNRs are symmetric) because for perfectly flat graphene the $\pi$-bonding is independent of angle in the plane. Third, the artificially-symmetric boundary conditions, together with the two-identical-atom bulk unit cell and single-orbital basis, produce mirror-image conduction- and valence-bands, for both bulk and AGNRs. Clearly the $p/d$ model overcomes these problems and provides a realistic passivation approach missing from the $p_z$-only model.

### III. APPLICATION: DEVICE CHARACTERISTICS

#### A. AGNR-MOSFET model

The differences between the two models are readily apparent when an AGNR is used as a nanodevice. We illustrate this point by modeling the AGNR-12 MOSFET described in Ref. 14. In Fig. 3 we graph the drain current $I_d$ of the MOSFET versus the gate voltage $V_{gs}$ for two different values of drain voltage. As in Ref. 14 the AGNR-12 is 82 cells (35 nm) long with a 15 nm channel between two 10 nm $N^+$ terminals. The $p_z$-only results are graphed with dotted lines and closed symbols while the *p/d* results are graphed with solid lines and open symbols, circles for $V_{ds} = 0.1\text{V}$ and squares for $V_{ds} = 0.5\text{V}$. As in Ref. 14 the $p_z$-only AGNR-12 is not passivated; the *p/d* AGNR-12 is hydrogen passivated as discussed above. Our $p_z$-only results differ slightly from those of Ref. 14 probably because they change the value of $V_{pp\pi}$ for the edge atoms and we do not.

The two sets of characteristics show significant differences. Although the ON currents in the two models agree well there are major differences in the OFF currents. At both drain biases the $p_z$-only model significantly underestimates the OFF current. This development is not surprising given its much larger gap as shown in Fig. 2 (a) and (c). In both models the OFF current is much larger at high drain bias. This behavior can be attributed to hole-induced barrier lowering (HIBL), whereby occupied states in the conduction band of the drain align with quasi-bound states in the channel valence band, allowing holes to tunnel into the channel[21]. Band edge graphs of our device suggest that this process is enhanced in the *p/d* model as compared to the $p_z$-only model. The *p/d* model indicates that the MOSFET will be significantly leakier than predicted by the $p_z$-only model, demonstrating the importance of accurate bandstructure models for transport simulations. Such a model is even more important when graphene nanoribbons are considered as band-to-band tunneling transistors[22], where the band gap is the crucial parameter.

The *p/d* model is also computationally efficient, though of course not as fast as the $p_z$-only model. Even though the *p/d* model increases the time to solve the AGNR-MOSFET quantum transport problem using a wave function approach equivalent to the Non-Equilibrium Green's Function (NEGF) formalism[23] by a factor ~4-5 as compared to the $p_z$-only model, this increase impacts the total computation time less than one would expect. The reason is that the bandstructure-independent Poisson equation solution and update of the self-consistent electrostatic potential consumes about 50% of the total time in the $p_z$ model. Overall, the total simulation time in the *p/d* model is about 3 times larger than in the $p_z$ model, but it does not exceed a couple of minutes on a single computer, guaranteeing a good balance between physical and computational efficiency.

### B. Rough AGNR Conductance

As discussed above (Sec. II), there is still some uncertainty surrounding AGNR gaps in a realistic device environment, so we examine the differences in the $p_z$-only and *p/d* models for an application which does not depend on the gap: the conductances of rough-edged AGNRs. (Because the transport is strictly intra-band the gap is irrelevant.) This issue is technologically significant because fabrication variances will generally produce AGNRs having rough edges, which will affect device performance. Line-edge roughness has been treated in the $p_z$-only model[12] and the third-nearest-neighbor $\pi$-bonded model[24] by adding an edge-disorder parameter as well as by actually removing carbon atoms from the nanoribbons in the $p_z$-only model.[25]  Here we examine the effects of line-edge roughness on AGNR differential conductance in the $p_z$-only and $p/d$ models.

We consider two test sets of rough AGNRs:  AGNR-12 and AGNR-13.  We simulate line-edge roughness by removing pairs of atoms from the edges with varying probabilities and calculate the resulting differential conductance, $G = dI_d/dV$ , using the forward-difference formula at $V = 0$ with $\Delta V = 10^{-5}$ eV.  As with the AGNR bands above, the $p_z$-only results are for unpassivated nanoribbons while in the $p/d$ model, after removal of carbon atoms the new edge carbons are hydrogen passivated as before.  For each nominal width we examine both $N$- and $P$-channel AGNRs, setting the Fermi level to the appropriate band-edge (conduction for $N$, valence for $P$).  The reported conductances are averages of 250 samples for each test case (model, fixed nominal width, and channel type).  The results are plotted in Figs. 4 (AGNR-12) and 5 (AGNR-13); note the logarithmic scale in Fig. 5.

In all cases we find significant differences between the two models.  In the AGNR-12 case (Fig. 4) the artificially-perfect symmetry for $N$- and $P$-channel AGNRs in the $p_z$-only model is readily apparent.  In contrast, the $p/d$ model displays clear differences in the two at any finite roughness.  This behavior is also present in the AGNR-13 case (Fig. 5), and is even visible on the logarithmic scale.  In both cases the conductance decreases more rapidly with roughness than in the AGNR-12 case, and in the $p/d$ model approaches a much lower limiting value.

The AGNR bands of Figs. 2(b)-(d) help qualitatively explain the differences in the various conductances.  When comparing the $p_z$-only bands for different width AGNRs, note that the bands are always centered around $E_p$. Removing carbons from an AGNR-12 results in a section of AGNR-11, which is either metallic ($p_z$-only model) or has a tiny (~0.2 eV) gap (*p/d* model). Wavefunction reflections at the discontinuities reduce the conductance, but the reductions are somewhat mitigated by the fact that in both models (and for both polarities) carriers are, roughly speaking, transmitting over a potential well. In the AGNR-13, carriers in the $p_z$-only model see a barrier in the AGNR-12 sections due to its larger gap, enhancing wavefunction reflections and leading to the generally higher conductances. In the *p/d* model, carriers do tunnel over potential wells as before, but there is a significant velocity mismatch which tends to enhance reflections due to the larger masses (both conduction and valence) in the AGNR-13 vs. the AGNR-12. Again, the generally higher conductances at the same roughness in the AGNR-12 case versus the AGNR-13 case are reasonable. The significant differences in the two models, viewed in light of the excellent agreement of the *p/d* approach with the DFT AGNR bands and corresponding failure of the $p_z$-only model to reproduce those bands, show that the *p/d* model should be much more accurate for calculating AGNR device characteristics and that the $p_z$-only model might well lead to incorrect conclusions about the performance of a device or the potential of a given device structure.

## IV. CONCLUSIONS

Motivated by the need for a nearest-neighbor approach which accurately models potential next-generation graphene nanodevices, we have introduced a six-band *p/d* model for the $\pi$-bands of graphene. Our model represents a significant improvement over the $p_z$-only model because it accurately reproduces the bulk graphene bands and provides a realistic hydrogen passivation approach. Our C-C parameters are fit to DFT+GW bands and the H-C parameters are fit to the LDA gaps in the AGNR series -7, -8, -9. Used with AGNRs from 5-13 (as well as 17, not shown) they give excellent agreement with LDA AGNR band gaps and shapes. We have employed these parameters to model the AGNR-12 MOSFET of Ref. 14, as well as the differential conductance of rough-edge AGNRs. In the case of the AGNR-12 MOSFET we find that at high drain bias the OFF current in the $p/d$ model is much higher than in the $p_z$-only model. In the rough AGNR simulations we find an artificial symmetry between $N$- and $P$-channel AGNRs in the $p_z$-only model, in contrast to clear polarity differences in the $p/d$ model. Furthermore the conductances in the two models differ considerably. Taken together these results demonstrate the importance of realistic bandstructure models for transport simulations.

## ACKNOWLEDGEMENTS

This work was partially supported by NSF grant EEC-0228390 that funds the Network for Computational Nanotechnology; by NSF PetaApps grant number 0749140; by NRI MIND; by NSF through TeraGrid resources provided by the National Institute for Computational Sciences (NICS); and by the New York State Interconnect Focus Center (XJ, NK, YZ, and SKN).

## APPENDIX

Here we demonstrate the superiority of the *p/d* model for bulk graphene using analytic band-edge energy expressions. For bulk, label the two carbon atoms in the graphene unit cell 'A' (on the lattice points) and 'B' (displaced by $\boldsymbol{\tau} = a\mathbf{e}_x$, $a = 0.142\text{nm}$), so that the Bloch states are:

$$\left| \omega C; \mathbf{k} \right\rangle = \frac{1}{\sqrt{N}} \sum_{j=1}^{N} \exp\left[ i\mathbf{k} \cdot \left( \mathbf{R}_j + \delta_{C,B} \boldsymbol{\tau} \right) \right] \left| \omega C; \mathbf{R}_j + \delta_{C,B} \boldsymbol{\tau} \right\rangle, \tag{1}$$

where $\omega \in \left\{ z, yz, zx \right\}$, $C \in \left\{ A, B \right\}$, $\mathbf{R}_j$ is a direct lattice vector, and $\mathbf{k}$ is a wavevector in the first Brillouin zone, both of which lie in the *x-y* plane. At $\Gamma$ the bands are purely *p*- or *d*-like, so that the valence- $\left( E_{-,\Gamma} \right)$ and conduction- $\left( E_{+,\Gamma} \right)$ bands have the same parameter dependence as in the nearest-neighbor $p_z$-only model,

$$E_{\pm,\Gamma} = E_p \mp 3V_{pp\pi}. \tag{2}$$

At *K* and *M* the Hamiltonian is most easily block-diagonalized by the transformations

$$\left| \omega \pm; \mathbf{k}_s \right\rangle = \frac{1}{\sqrt{2}} \left[ \left| \omega A; \mathbf{k}_s \right\rangle \pm \exp\left( i\varphi_s \right) \left| \omega B; \mathbf{k}_s \right\rangle \right], s \in \left\{ K, M \right\}, \tag{3}$$

where $\mathbf{k}_s$ is the wavevector at the symmetry point, *s*, and $\varphi_M = \pi/3$, $\varphi_K = -2\pi/3$. At *M* only the *z* and *zx* states are coupled and the Hamiltonian breaks into two two-dimensional subspaces, $\left\{ \left| z\pm; \mathbf{k}_M \right\rangle, \left| zx \mp; \mathbf{k}_M \right\rangle \right\}$, whence come the highest valence band, $E_{-,M}$ (upper sign pair), and lowest conduction band, $E_{+,M}$ (lower sign pair). These energies are:

$$E_{\pm,M} = \frac{E_p + E_d \pm \left( U_{zx} - V_{pp\pi} \right)}{2} - \frac{1}{2} \sqrt{\left[ \left( E_d - E_p \right) \pm \left( U_{zx} + V_{pp\pi} \right) \right]^2 + 16V_{pd\pi}^2} \tag{4}$$

In eq. (4) the parameter $U_{zx} = \left( 3V_{dd\delta} - V_{dd\pi} \right)/2$. At *K* all six states are coupled, but the Hamiltonian breaks into two three-dimensional subspaces:

$\left\{\left|z\pm;\mathbf{k}_K\right\rangle,\left|yz\pm;\mathbf{k}_K\right\rangle,\left|zx\mp;\mathbf{k}_K\right\rangle\right\}$, whose characteristic polynomials factor into a linear and a quadratic term, the latter identical for the two subspaces. The lower root of the common quadratic term gives the low-energy $K$ degeneracy:

$$E_K = \frac{E_p + E_d}{2} - \frac{1}{2}\sqrt{\left(E_d - E_p\right)^2 + 18V_{pd\pi}^2} \ . \tag{5}$$

Equations (4)-(5) show that in the *p/d* model the conduction- and valence-bands at *M* generally are *not* symmetric about *K*, unlike those of the $p_z$-only model, as can be seen by setting $V_{pd\pi} = 0$ in eqs. (4)-(5). As discussed above, the ability to reproduce this asymmetry leads to a superior agreement with the DFT+GW bands. In particular, note that the $p_z$-only model significantly overshoots the conduction-band at *M* (Fig. 1). Because bulk states in the vicinity of *K* and on the *K-M* line play a major role in the bands of AGNRs, this deficiency of the customary model can lead to poor AGNR bands as already seen in Fig. 2.

**Table I:**  C-C and H-C onsite and nearest-neighbor tight-binding parameters used in the *p/d* model; all values are in eV.  To simplify the treatment we employ only a single H-C *pd* nearest-neighbor parameter:  $p(H)d(C)\pi = p(C)d(H)\pi$.

| C-C | | H-C | |
|---|---|---|---|
| $E_p(C)$ | 1.2057 | $E_p(H)$ | 13.04020 |
| $E_d(C)$ | 24.1657 | $E_d(H)$ | 20.9020 |
| $pp\pi$ | -3.2600 | $pp\pi$ | -0.61754 |
| $pd\pi$ | 2.4000 | $pd\pi$ | 3.41170 |
| $dd\pi$ | 3.6000 | $dd\pi$ | 10.44660 |
| $dd\delta$ | -7.4000 | $dd\delta$ | -13.96340 |

Figure Captions

**Figure 1:** **(a)** Bulk bands of graphene from the DFT+GW calculations used to fit the parameters of Table I(a) (see text). The highlighted area around the $K$-point is expanded in part (b). **(b)** the bulk bands of all three models in the vicinity of $K$: DFT+GW (diamonds), the *p/d* model (solid lines) and the $p_z$-only model (dotted lines). The $p_z$-only model has parameters $E_p = 0.12742\text{eV}, V_{pp\pi} = -2.7\text{eV}$; the nonzero onsite term is chosen to align its $K$-point with that of the other calculations and the common $K$-point energy is indicated by the heavy dashed horizontal line labeled $E_K$. Note the asymmetry of the DFT+GW and *p/d* bands about $E_K$ in contrast to the exact symmetry of the $p_z$-only bands about this energy.

**Figure 2:** Gaps (a) and bands (b)-(d) of AGNRs; the DFT and *p/d* model AGNRs are hydrogen passivated while the $p_z$-only model AGNRs are not (see text), and the uppermost valence band of each model is aligned to facilitate comparison. In panel (a) the DFT gaps are plotted with diamonds, those of the *p/d* model with open circles, and those of the $p_z$-only model with squares. Lines, which are guides to the eye, denote families: $3n+1$ (dashed), $3n$ (solid), and $3n+2$ (dotted). The hydrogen and H-C parameters of Table I were optimized to the DFT results for only AGNR-7, -8, and -9. All other AGNRs use these same parameters. Note the excellent agreement of the *p/d* model with DFT and the poor agreement of the $p_z$-only results.

**Figure 3:** Current-voltage characteristics of the AGNR-12 MOSFET of Ref. 14 as calculated with the $p_z$-only (dotted lines and closed symbols) and *p/d* models (solid lines and open symbols) for two different drain biases.  Computed points are indicated by symbols:  circles $\left(V_{ds}=0.1\text{V}\right)$ and squares $\left(V_{ds}=0.5\text{V}\right)$.  Note in Fig. 2(c) the larger AGNR bandgap in the $p_z$-only model versus the *p/d* model.

**Figure 4:** Differential conductance of a rough AGNR-12 versus line edge roughness probability (see text).  Symbols are calculated conductances:  Open circles ($p_z$, *N* channel), solid circles ($p_z$, *P* channel), open diamonds (*p/d*, *N* channel), or solid diamonds (*p/d*, *P* channel).  Each computed point is the average of 250 samples.  Note the artificial symmetry of the *N*- and *P*-channel AGNRs in the $p_z$-only model which does not appear in the *p/d* model.

**Figure 5:** Differential conductance of a rough AGNR-13 versus line edge roughness probability (see text); note the logarithmic scale.  Symbols are calculated conductances: Open circles ($p_z$, *N* channel), solid circles ($p_z$, *P* channel), open diamonds (*p/d*, *N* channel), or solid diamonds (*p/d*, *P* channel).  Each computed point is the average of 250 samples.  Note the artificial symmetry of the *N*- and *P*-channel AGNRs in the $p_z$-only model which does not appear in the *p/d* model.

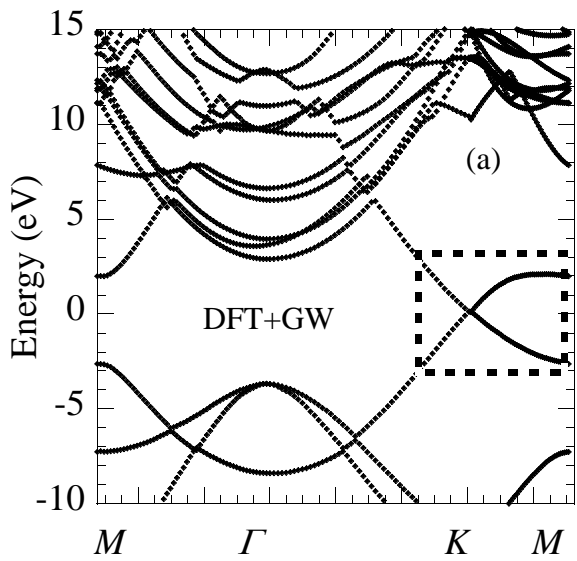
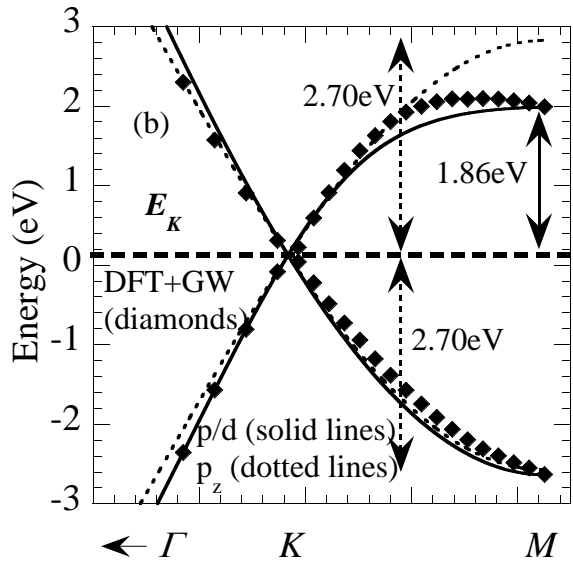

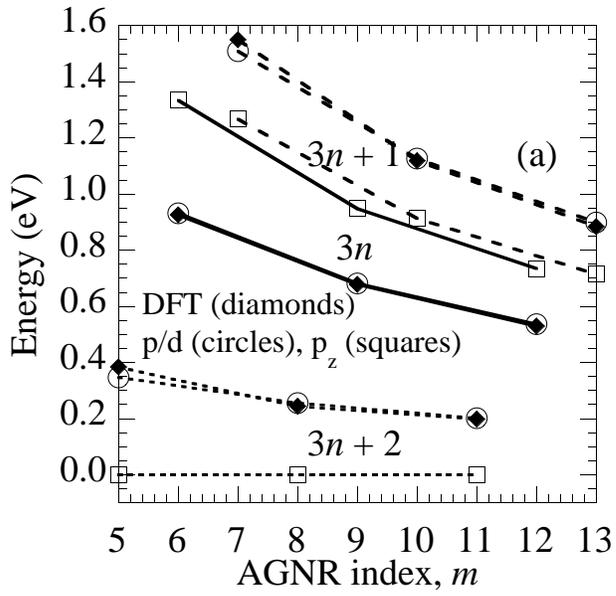

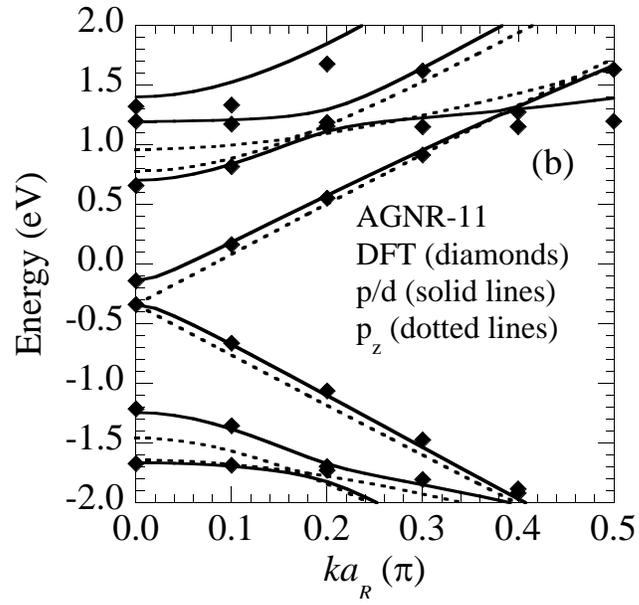

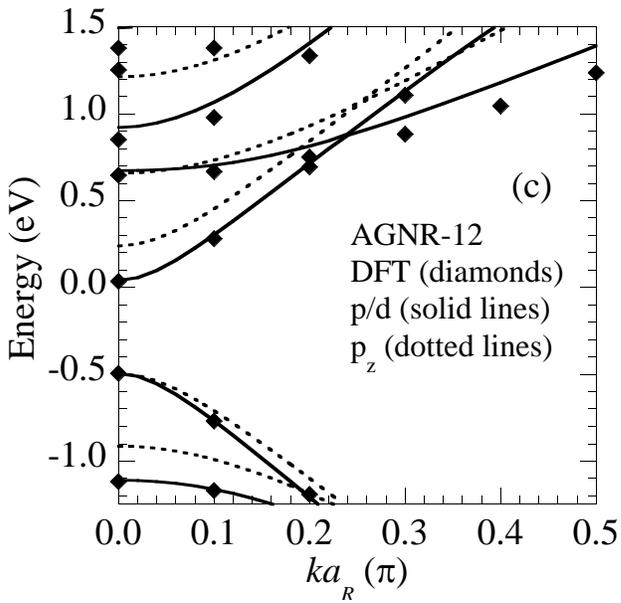

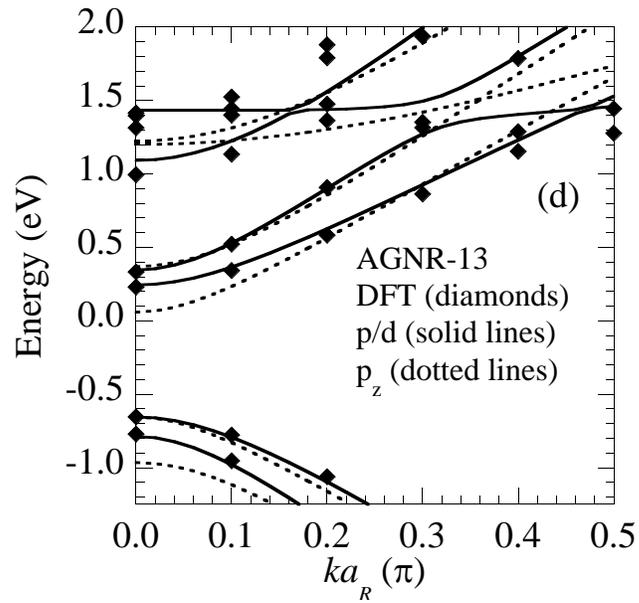

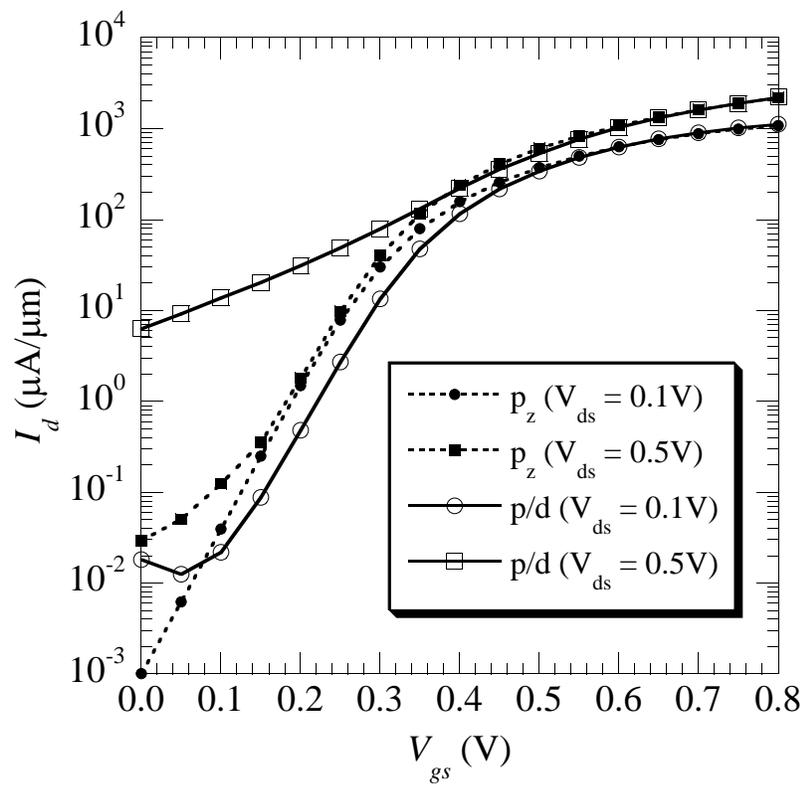

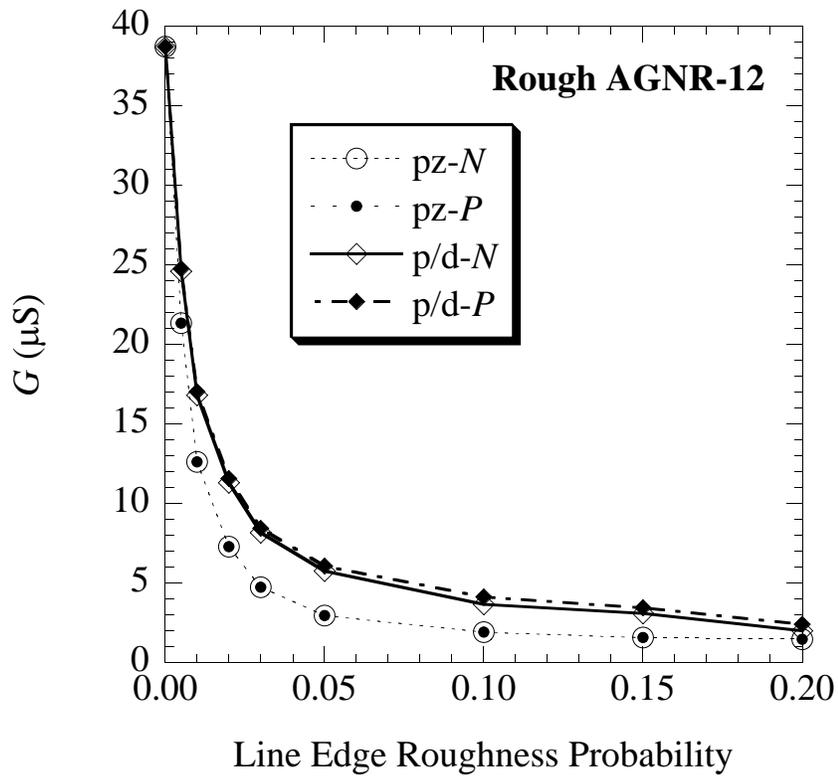

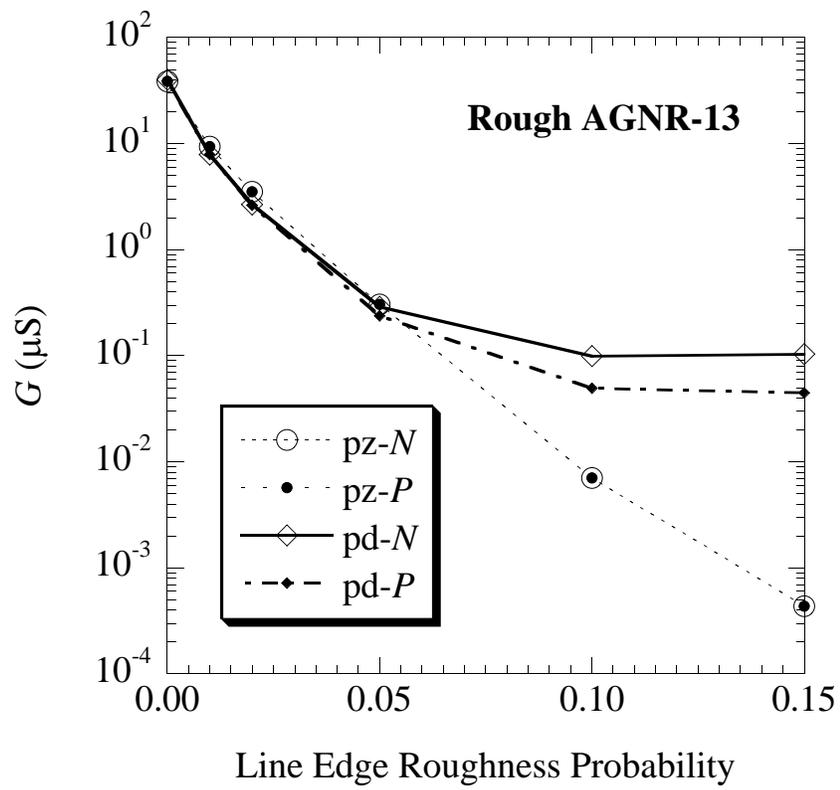

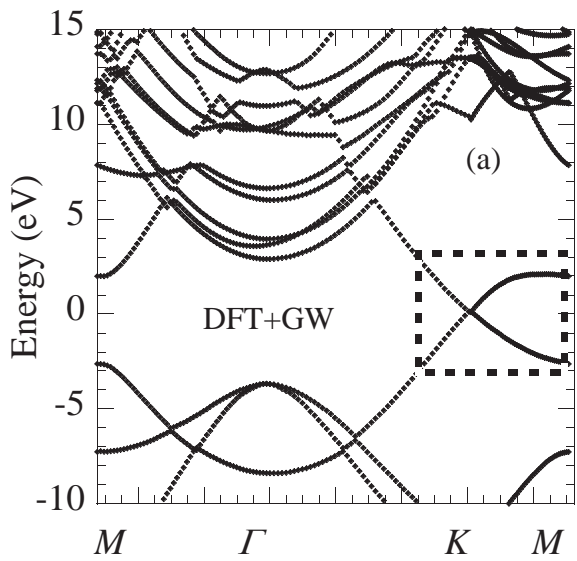
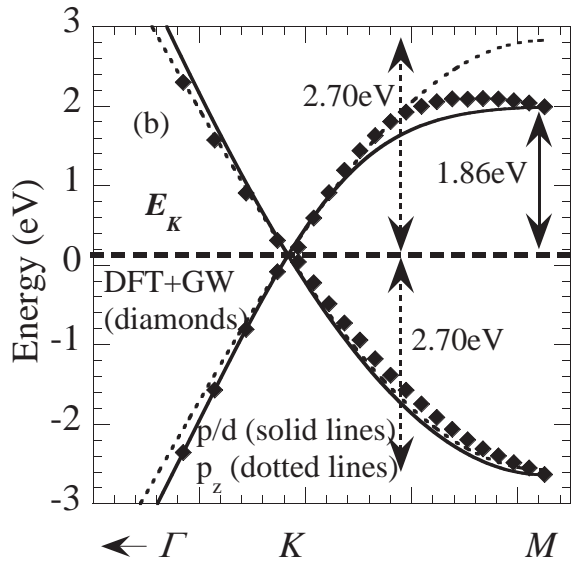

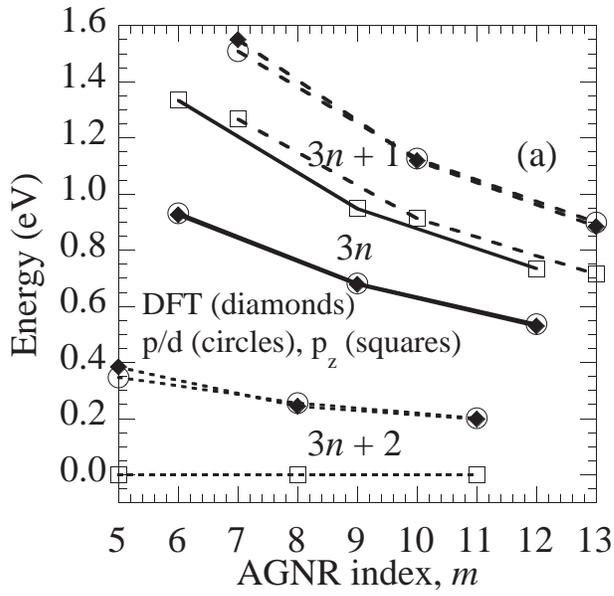

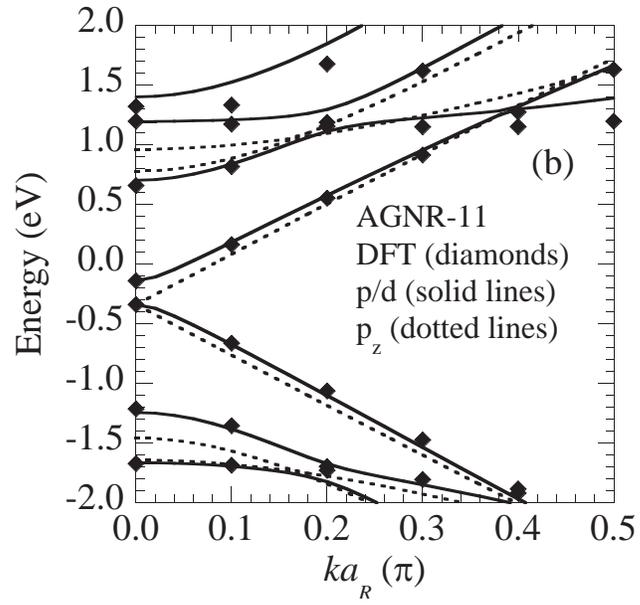

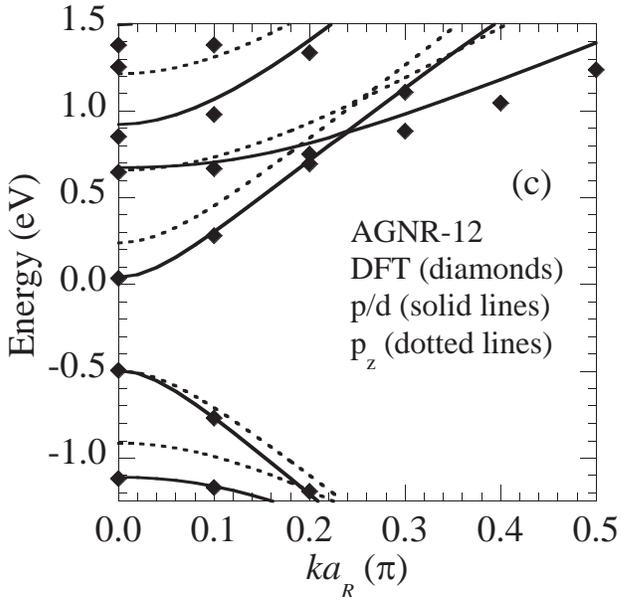

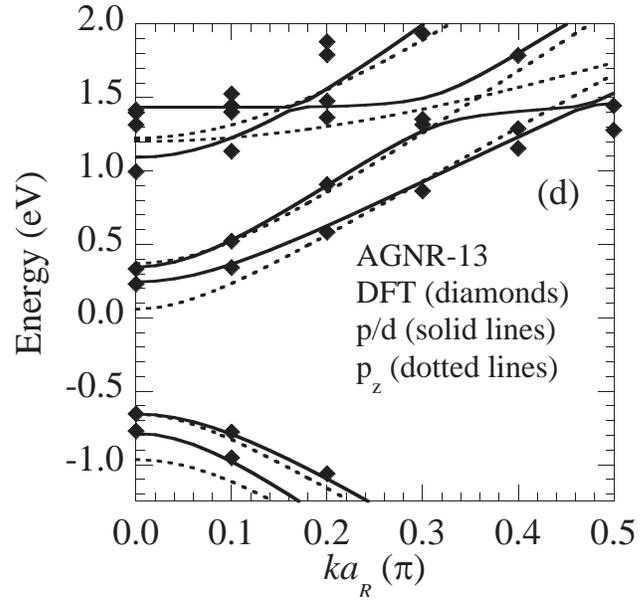

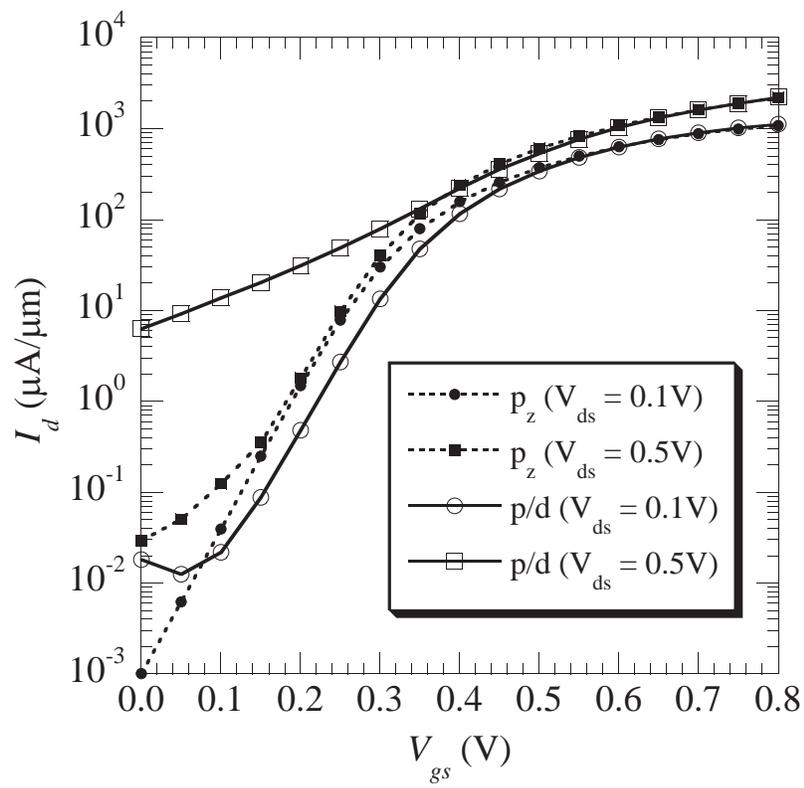

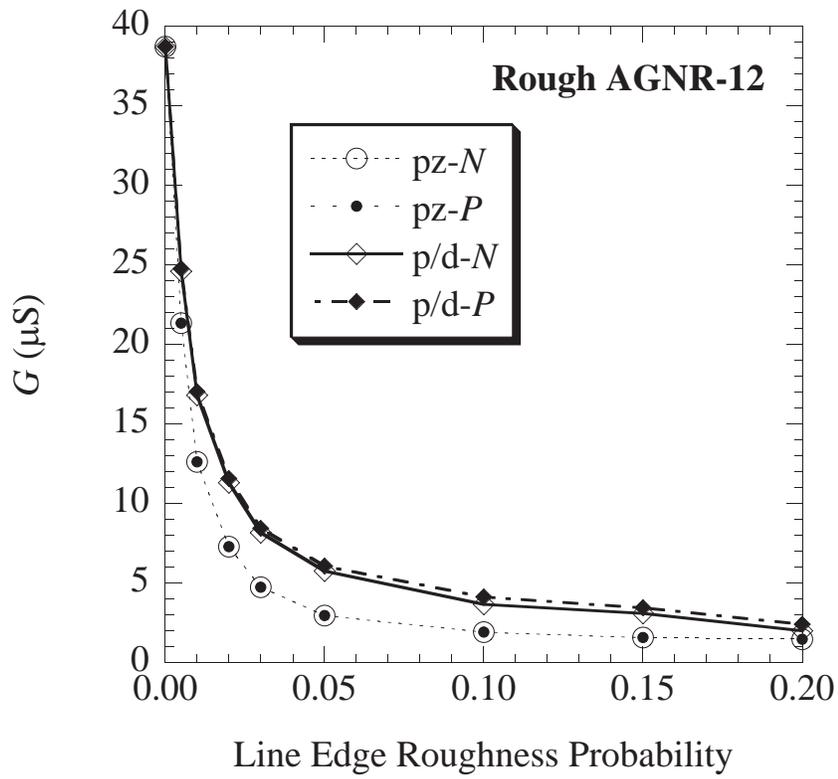

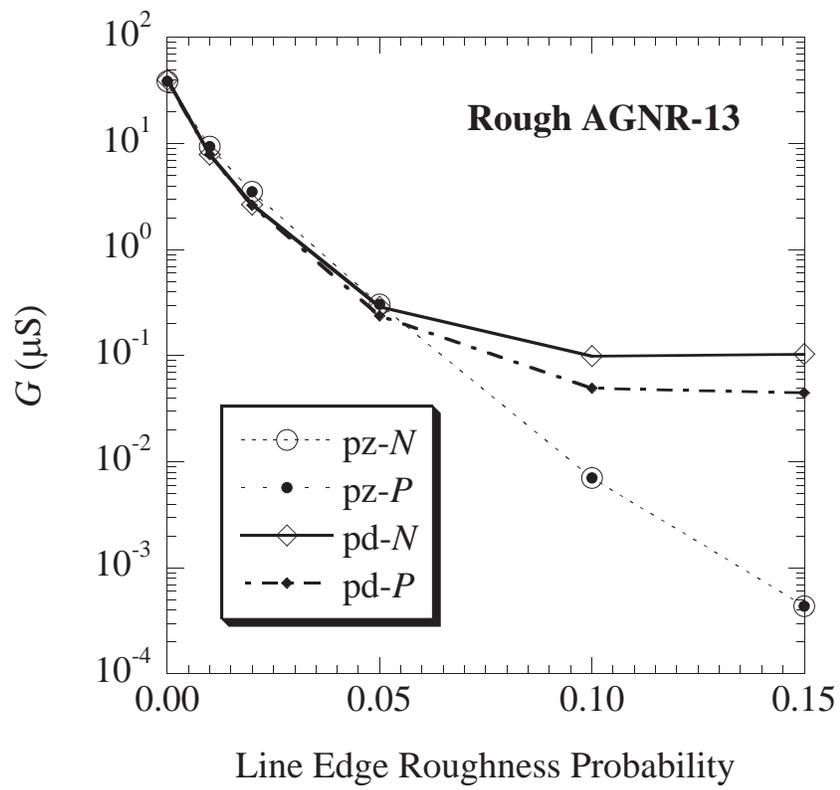